\ifx\pdfoutput\undefined        
  \documentclass[twocolumn]{article}
  \usepackage{url}
\else                           
  \documentclass[pdflatex]{article}
  
  \usepackage[bookmarks=false]{hyperref}
\fi

\usepackage{graphicx}
\usepackage{url}
\newcommand{\EQ}{\begin{equation}}
\newcommand{\EN}{\end{equation}}
\newcommand{\EQA}{\begin{eqnarray}}
\newcommand{\ENA}{\end{eqnarray}}
\newcommand{\eq}[1]{(\ref{#1})}
\newcommand{\EEq}[1]{Equation~(\ref{#1})}
\newcommand{\Eq}[1]{Eq.~(\ref{#1})}
\newcommand{\Eqs}[2]{Eqs.~(\ref{#1}) and~(\ref{#2})}
\newcommand{\EEqs}[2]{Equations~(\ref{#1}) and~(\ref{#2})}

\newcommand{\Eqss}[2]{Eqs.~(\ref{#1})--(\ref{#2})}
\newcommand{\eqss}[2]{(\ref{#1})--(\ref{#2})}

\newcommand{\Fig}[1]{Fig.~\ref{#1}}

\newcommand{\Figss}[2]{Figs~\ref{#1}--\ref{#2}}

\newcommand{\bra}[1]{\langle #1\rangle}

\newcommand{\xx}{\mbox{\boldmath $x$} {}}

\newcommand{\uu}{\mbox{\boldmath $u$} {}}

\newcommand{\nab}{\mbox{\boldmath $\nabla$} {}}

\newcommand{\DDD}{{\cal D} {}}

\newcommand{\dd}{{\rm d} {}}

\newcommand{\ypnas}[5]{ (#1) #5, {\em Proc.\ Natl.\ Acad.\ Sci.\ }{\bf #2}, #3--#4.}

\newcommand{\ysci}[5]{ (#1) #5, {\em Science }{\bf #2}, #3--#4.}

\newcommand{\ynat}[5]{ (#1) #5, {\em Nature }{\bf #2}, #3--#4.}

\newcommand{\ypreN}[4]{ (#1) #4, {\em Phys.\ Rev.\ E }{\bf #2}, #3.}

\newcommand{\yptrs}[5]{ (#1) #5, {\em Phil.\ Trans.\ Roy.\ Soc.\ }{\bf #2}, #3--#4.}

\newcommand{\yapj}[5]{ (#1) #5, {\em Astrophys.\ J.\ }{\bf #2}, #3--#4.}

\newcommand{\ypf}[5]{ (#1) #5, {\em Phys.\ Fluids }{\bf #2}, #3--#4.}

\newcommand{\yphy}[5]{ (#1) #5, {\em Physica } {\bf #2}, #3--#4.}

\newcommand{\yoleb}[5]{ (#1) #5, {\em Orig.\ Life Evol.\ Biosph.\ }{\bf #2}, #3--#4.}
\newcommand{\yab}[5]{ (#1) #5, {\em Astrobiol.\ }{\bf #2}, #3--#4.}
\newcommand{\yjour}[6]{ (#1) #6, {\em #2} {\bf #3}, #4--#5.}
\newcommand{\yjourS}[6]{ (#1) #6 {\em #2} {\bf #3}, #4--#5.}

\newcommand{\ybook}[3]{ (#1) {\em #2}, #3.}

\newcommand{\sjour}[3]{ (#1) #3, {\em #2} (submitted).}

\newcommand{\poleb}[2]{ (#1) #2, {\em Orig.\ Life Evol.\ Biosph.} (in press).}
\newcommand{\ea}{{\em et al., }}
\newcommand{\eaa}{{\em et al. }}
\def\half{{\textstyle{1\over2}}}

\def\onethird{{\textstyle{1\over3}}}

\def\quarter{{\textstyle{1\over4}}}

\newcommand{\s}{\,{\rm s}}

\newcommand{\mtwopers}{\,{\rm m^2/s}}

\newcommand{\cm}{\,{\rm cm}}
\newcommand{\cms}{\,{\rm cm/s}}

\newcommand{\km}{\,{\rm km}}
\newcommand{\kmyr}{\,{\rm km/yr}}
\newcommand{\kmtwoperyr}{\,{\rm km^2/yr}}

\newcommand{\yr}{\,{\rm yr}}

\graphicspath{{./fig/}{./png/}}

\begin{document}

\title{How Long can Left and Right Handed Life Forms Coexist?}
\author{Axel Brandenburg and Tuomas Multam\"aki \\
NORDITA, Blegdamsvej 17, DK-2100 Copenhagen \O, Denmark
}

\date{
Draft version, \today,~ $ $Revision: 1.90 $ $}

\maketitle

\abstract{\bf
Reaction-diffusion equations based on a polymerization model
are solved to simulate the spreading of hypothetic left and right
handed life forms on the Earth's surface.
The equations exhibit front-like behavior as is familiar from
the theory of the spreading of epidemics.
It is shown that the relevant time scale for achieving global
homochirality is not, however, the time scale of front propagation,
but the much longer global diffusion time.
The process can be sped up by turbulence and large scale flows.
It is speculated that, if the deep layers of the early ocean
were sufficiently quiescent, there may have been the
possibility of competing early life forms with opposite handedness.
}

\section*{Introduction}

There has been a remarkable development in assessing the probability
of extraterrestrial life in the Universe.
Although higher life forms are now believed to be exceedingly rare,
primitive life forms may be more wide spread than what has been
believed some 20 years ago (Ward and Brownlee, 2000).
Two major discoveries have contributed to this view:
(i) microfossil and carbon isotope evidence
that life has existed on Earth at least as early
as 3.8 billion years ago (Schopf, 1993; Mojzsis \ea 1996),
and (ii) the discovery of extremophilic
life forms on Earth making it plausible that life can exist (or has
existed) on other celestial bodies in our solar system and beyond.
However, Cleaves and Chalmers (2004) present a more cautions view
quoting evidence that some of the oldest viable organisms were actually
only mesophiles, and not at least hyperthermophiles.

Life may either have originated elsewhere and then delivered to Earth
[but again, see Cleaves and Chalmers (2004) for a more cautions view],
or it may have been created locally from prebiotic chemistry,
e.g.\ in hydrothermal systems at the ocean floor
(Russell and Hall, 1997; Martin and Russell, 2003).
In either case there is the possibility of multiple sites on Earth,
or alternatively in space, where different life forms may have
coexisted for some time --
even though on Earth only one life form has prevailed in the end.
Indeed, the DNA of the most primitive life forms on Earth
suggests that it is based
on a single common ancestor (e.g., Collins \ea 2003).
One important characteristics of life, as we know it, is the handedness
(or chirality) of amino acids and sugars.
It is well known that
the sugars generated in living organisms are dextrorotatory.
Examples include the glucose produced in plants via photosynthesis,
but also the ribose sugars in the backbones of the nucleotides
of the DNA and RNA.
Conversely, the amino acids in proteins of living organisms are all
left-handed.

A commonly discussed pathway for the origin of life is a pre-RNA world
(Bada, 1995; Nelson \ea 2000), that could have been -- in its earliest
form -- achiral.
An early peptide world composed of glycines has been discussed
(Milner-White and Russell, 2004) and also
peptide nucleic acids (PNA) based on a glycine backbone
are achiral (e.g., Pooga \ea 2001).
PNA molecules are frequently discussed for transcribing the
genetic code of early life forms (Nielsen \ea 1993) and even artificial
life forms (Rasmussen \ea 2003).
However, PNA based on other amino acids is chiral (Tedeschi \ea 2002),
and it might therefore be plausible that the onset of chirality could
have emerged during the evolution of the PNA world and hence
after the onset of life.

There is a priori no reason why life on Earth could not have been
is based on nucleotides with left-handed sugars that catalyze the production
of right-handed amino acids, i.e.\ just opposite to what it actually is.
It is now widely accepted that at some point around the time when the
first life has emerged (or some time thereafter)
there must have been a random selection favoring
one particular chirality.
Whether it was right or left handed must have been a matter of chance
and can be explained by self-catalytic polymerization of nucleotides
(Frank, 1953).
An important aspect of this process is the cross-inhibition by
nucleotides with the opposite chirality which would spoil the polymer
and prevent further polymerization on the corresponding end of the
polymer (Joyce \ea 1984).
Although polymerization of alternating left and right handed nucleotides
is usually impossible (Kozlov \ea 1998), this can be different for amino
acids in more complicated molecules, for example certain siderophores
(``iron bearers'') can contain mixed left and right handed amino acids
(Martinez \ea 2000).
Another example are the so-called ``nests'', i.e.\ are pairs of
left and right handed amino acids within relatively short sequences
(Watson and Milner-White, 2002; Pal \ea 2002).

The first detailed polymerization model taking such cross-inhibition
into account was proposed by Sandars (2003).
This model was studied further by Brandenburg \eaa (2004a, hereafter
referred to as BAHN) and Wattis and Coveney (2004).
Since the influential paper of Frank (1953) the various models
that have been devised
to explain the origin of the same chirality (or homochirality) of
nucleotides and amino acids were all based on a pair of two ordinary differential
equations describing the time evolution of some collective properties
of the two types of polymers with left and right handed building blocks
(e.g., Saito and Hyuga, 2004a; hereafter referred to as SH).
As was shown in the paper of BAHN, a modified version of the model of
SH can actually
be derived as a reduction of the full polymerization model of Sandars (2003).

In the present paper we extend the model of BAHN by including
the fact that opposite chiralities are likely to originate at different
places.
At those places where life forms of opposite chirality meet, one of
the two will eventually dominate and extinguish the other.
If the time for this to happen was sufficiently long, life
forms of opposite handedness may have coexisted for some time at different places
on the early Earth.

Mathematically speaking, our model falls in the class of diffusion-reaction
(or advection-diffusion-reaction) equations that are being discussed in many
different fields; see the book by Murray (1993) for biological applications.
In this paper the reactions correspond to the auto-catalytic polymerization
which corresponds to a local instability with a given linear growth rate.
The importance of spatial extent has already been emphasized by
Saito and Hyuga (2004b) who generalized the SH model
by using a Monte-Carlo method as it is used in percolation studies.

We begin by outlining the basic alterations to models without spatial
extent and discuss relevant time scales for achieving homochirality
in a spatially extended system.
We then briefly review the polymerization model of Sandars (2003) and
the reduction of BAHN, discuss the effects of spatial extent and
present results of simulations.


\section*{Relevant time scales}

Since the seminal paper by Frank (1953) we know that,
in the case without spatial extent,
an infinitesimally small initial
enantiomeric excess $\eta$ (defined in the range $-1\leq\eta\leq1$; see below)
can grow exponentially like
\EQ
\eta=\eta_0 e^{\lambda t}\quad
\left(\mbox{for $|\eta-\eta_0|\ll1$}\right).
\EN
Here $\eta_0$ is the initial value, which must be close to the
racemic value of zero (see BAHN for the calculation of $\lambda$).
The growth rate depends on the reaction coefficients
and the source term replenishing the substrate, from which new
monomers can grow.
In the case with spatial extent, however, there is the possibility that
life forms with opposite handedness can emerge independently at different sites.
This can substantially delay the evolution toward homochirality, because
now the relevant time scale is governed by the time it takes to extinguish
already existing life forms of opposite handedness.

The situation is somewhat analogous to the propagation of epidemics
whose expansion over the Earth's surface is known to be related to
the speed at which fronts propagate in reaction-diffusion systems.
A prototype of such a system is the Fisher equation
(Fisher, 1937; Kaliappan, 1984; Murray, 1993)
\EQ
{\partial c\over\partial t}=\lambda c\left(1-{c\over c_0}\right)
+\kappa{\partial^2 c\over\partial x^2},
\label{FisherEqn}
\EN
where $c$ is the concentration, $\lambda$ is the growth rate in the
homogeneous limit ($c$ independent of $x$), $c_0$ is a saturation
amplitude, and $\kappa$ is a diffusion coefficient.
\EEq{FisherEqn} admits propagating wave solutions of the form
$c=c(x-v_{\rm front}t)$ where
\EQ
v_{\rm front}\approx2(\kappa\lambda)^{1/2}
\label{vfront}
\EN
is the speed of the front [Murray (1993); see also M\'endez \eaa (2003)
for the case of a heterogeneous medium].
A typical application of reaction-diffusion equations
is the spreading of the Black Death in Europe in
the fourteenth century.
In this connection, Murray (1993) quotes
an effective diffusion coefficient of about $3\times10^{4}\kmtwoperyr$
(corresponding to $10^{3}\mtwopers$) and a reaction rate
(or in this case a `mortality rate') of $15\yr^{-1}$
(corresponding to $\lambda\sim5\times10^{-7}\s^{-1}$), which
gives $v_{\rm front}\approx5\cms$ or about $1300\kmyr$.

Applying the same front speed ($1300\kmyr$)
to the homochiralization over
the whole Earth surface (typical scale $L\sim10\,000\km$),
one finds a typical time scale $\tau_{\rm front}=L/v_{\rm front}$,
which would be on the order of 10 years.
However, it will turn out that the time scale involving front
propagation is too optimistic an estimate for the problem at hand,
because once right and left handed
life forms meet, the reaction front will almost come to a halt.
As it turns out, the only reason why these fronts can still propagate
is because of their curvature.
This leads eventually to the complete dominance of only one handedness.
The time scale for this to happen turns out to be
the global diffusion time scale
\EQ
\tau_{\rm diff}=L^2/\kappa.
\EN
Putting in the same numbers as above one finds $\sim3000\yr$.
This is still rather short, but this is because the adopted value
for $\kappa$ is more like an effective (eddy) diffusivity and
hence rather large compared with the usual molecular values
quoted by Cotterill (2002) for air ($\kappa=10^{-5}\mtwopers$)
or water ($\kappa=10^{-9}\mtwopers$).
These values would lead to diffusion times in excess of the age of
the universe.

Larger values of $\kappa$ are probably more realistic in that
they take into account the effects of additional macroscopic
mixing processes that are required to establish global homochirality.
Examples include ocean currents, surface waves, as well as winds
and turbulence in the atmospheric boundary layer (e.g.\ Garrett, 2003).
Below we shall refer to the corresponding time scale as the
turbulent time scale
\EQ
\tau_{\rm turb}=L/u_{\rm rms},
\EN
where $u_{\rm rms}=\bra{\uu^2}^{1/2}$ is the root mean square velocity.

A commonly used procedure to estimate turbulent diffusion times is in terms
of enhanced turbulent (eddy) diffusivities.
We return to this in more detail, but for now we just note that
in the ocean values between $\sim 10^{-4}\mtwopers$ for the
vertical eddy diffusivity and up to $\sim 10^{3}\mtwopers$ for the
horizontal eddy diffusivity are typical (see Chap.~8.5 of Stewart, 2003).
This would again suggest turbulent diffusion times of around 3000 years,
which would then be the time required for achieving global homochirality
in the ocean.

The estimates outlined above can be put on a more quantitative basis.
Before we can study the analogous equation to the Fisher equation,
we briefly review the basis of the polymerization model of Sandars (2003)
and its reductions as proposed by BAHN.
Since we are interested in the spatially extended case,
we add the effects of diffusion and advection to these equations.
We therefore consider the case where the concentrations of the different
polymers $[L_n]$ and $[R_n]$, obey advection-diffusion-reaction equations, 
i.e.\ everywhere in the equations in the paper by BAHN we replace
\EQ
{\dd\over\dd t}\rightarrow
{\DDD\over\DDD t}\equiv{\partial\over\partial t}
+\uu\cdot\nab-{\kappa}\nabla^2.
\EN
Here, $\uu$ and $\kappa$ are the velocity vector
and the diffusion coefficient, respectively.

\section*{The polymerization model}

In the polymerization model of Sandars (2003) it is assumed that a
polymer $L_n$ with $n$ left-handed building blocks can grow on either
end by a left or right handed monomer, $L_1$ or $R_1$, respectively,
with a reaction coefficient $2k_S$ or $2k_I$, respectively.
The latter reaction is referred to as enantiomeric cross-inhibition
leading to a polymer whose one end is now `spoiled', allowing further
growth only on the remaining other end at rates $k_S$ or $k_I$
for a reaction with $L_1$ or $R_1$, respectively.
(The left handed polymers $L_n$ are not to be confused with the
size of the domain $L$.)
The full set of reactions included in the model of Sandars (2003) is thus
\begin{eqnarray}
L_{n}+L_1&\stackrel{2k_S~}{\longrightarrow}&L_{n+1},
\label{react1}\\
L_{n}+R_1&\stackrel{2k_I~}{\longrightarrow}&L_nR_1,
\label{LnR1react}
\label{react2}\\
L_1+L_{n}R_1&\stackrel{k_S~}{\longrightarrow}&L_{n+1}R_1,
\label{react3}\\
R_1+L_{n}R_1&\stackrel{k_I~}{\longrightarrow}&R_1L_nR_1,
\label{react4}
\end{eqnarray}
where $k_S$ and $k_I$ are the reaction coefficients for monomers
to a polymer of the same or of the opposite handedness, respectively.
For all four equations we have the complementary reactions
obtained by exchanging $L\rightleftharpoons R$.

In addition, new monomers of either chirality are being regenerated
from a substrate $S$ at a rate that may depend on
the already existing excess of one chirality (i.e.\ the enantiomeric
excess), so
\EQ
S\stackrel{k_C C_R~}{\longrightarrow} R_1,\quad
S\stackrel{k_C C_L~}{\longrightarrow} L_1,\quad
\EN
where $k_C$ is proportional to the growth rate of monomers and
$C_R$ and $C_L$ determine the enzymatic enhancement of right and
left handed monomers.
Sandars (2003) assumed that $C_R=R_N$ and $C_L=L_N$, but different
alternative proposals have been made, for example
$C_R=\sum R_n$ and $C_L=\sum L_n$ (Wattis and Coveney 2004) or
$C_R=\sum nR_n$ and $C_L=\sum nL_n$ (BAHN).
The overall behavior of the model is not however affected by these
different choices.

The governing equations, ignoring spatial dependencies, are
(for $n\ge3$)
\begin{equation}
{\dd[L_{n}]\over\dd t}=2k_S[L_1]\Big([L_{n-1}]-[L_{n}]\Big)
-2k_I[L_{n}][R_1],
\label{Ln_new2}
\end{equation}
\begin{equation}
{\dd[R_{n}]\over\dd t}=2k_S[R_1]\Big([R_{n-1}]-[R_{n}]\Big)
-2k_I[R_{n}][L_1],
\label{Rn_new2}
\end{equation}
where the factor 2 in front of $k_S$ and $k_I$ reflects the fact that
the monomers can attach to either side of the polymer.
For $n=2$, however, $m$ identical monomers interact, and the number of
corresponding pairs is not $m^2$, as for non-identical pairs ($n>2$),
but only $\half m(m-1)\approx\half m^2$,
canceling the factor 2 in front of the $[L_1]^2$ term
(see Sandars 2003, and also BAHN). Thus, we have
\begin{equation}
{\dd[L_{2}]\over\dd t}=k_S[L_1]^2-2k_S[L_1][L_{2}]
-2k_I[L_{2}][R_1],
\label{Ln_new3}
\end{equation}
\begin{equation}
{\dd[R_{2}]\over\dd t}=k_S[R_1]^2-2k_S[R_1][R_{2}]
-2k_I[R_{2}][L_1].
\label{Rn_new3}
\end{equation}
The evolution equations for $[L_1]$ and $[R_1]$ are given by
\begin{equation}
{\dd[L_1]\over\dd t}=Q_L-\lambda_L[L_1],\quad
{\dd[R_1]\over\dd t}=Q_R-\lambda_R[R_1],
\label{L1R1}
\end{equation}
where
\begin{eqnarray}
\lambda_L=
2k_S\sum_{n=1}^{N-1}[L_n]
+2k_I\sum_{n=1}^{\tilde{N}}[R_n]
\nonumber \\
+k_S\sum_{n=2}^{N-1}[L_nR]
+k_I\sum_{n=2}^{\tilde{N}}[R_nL],
\label{lambdaL}
\end{eqnarray}
and
\begin{eqnarray}
\lambda_R=
2k_S\sum_{n=1}^{N-1}[R_n]
+2k_I\sum_{n=1}^{\tilde{N}}[L_n]
\nonumber \\
+k_S\sum_{n=2}^{N-1}[R_nL]
+k_I\sum_{n=2}^{\tilde{N}}[L_nR],
\label{lambdaR}
\end{eqnarray}
are the decay rates that quantify
the losses associated with the reactions
\eqss{react1}{react4}, respectively.
Here $\tilde{N}=N-1$ in the model of Sandars (2003) and
$\tilde{N}=N$ in the model of BAHN.

New monomers are generated from a substrate with concentration $[S]$ via
the source terms $Q_L$ and $Q_R$,
\EQA
Q_L=k_C[S]\left(pC_L+qC_R\right),\\
Q_R=k_C[S]\left(pC_R+qC_L\right),
\label{QLRdef}
\ENA
where $k_C$ is the corresponding reaction coefficient,
$p=\half(1+f)$ and $q=\half(1-f)$, and $f$ is the
fidelity of the enzymatic reactions.
The concentration of the substrate, in turn, is maintained
by a source $Q$, via
\EQ
{\dd[S]\over\dd t}=Q-\left(Q_L+Q_R\right),
\EN
where $Q_L+Q_R=k_C[S](C_L+C_R)$; see \Eq{QLRdef}.
In general, we expect $C_L$ and $C_R$ to be functions of
$L_n$ and $R_n$, respectively.
Sandars (2003) assumed $C_L=[L_N]$ and $C_R=[R_N]$, i.e.\ the
catalytic effect depends on the concentrations of the longest possible
chains of left and right handed polymers, but other choices are possible
(see BAHN).
The degree of handedness (or enantiomeric excess) is
generally characterized by the ratio of the difference to the
sum of the number of left and right handed building blocks, i.e.\
\EQ
\eta={E_R-E_L\over E_R+E_L}
\label{etadefn}
\EN
where $E_R=\sum n[R_n]$ and $E_L=\sum n[L_n]$ measure the total number
of left and right handed building blocks.

The important point is that the polymerization equations show bifurcation
behavior where $f$ is the control parameter and $\eta$ is the control
parameter.
There is a critical value, $f_{\rm crit}$, above which the racemic solution
($\eta=0$) is linearly unstable (with a growth rate 
$\dd\ln\eta/\dd t\equiv{\lambda}$) and a new solution
with enantiomeric excess is attained.
Only for $f=1$ is the enantiomeric excess complete, i.e.\ $\eta=1$ or $-1$.
The value of $f_{\rm crit}$ depends on the ratio $k_I/k_S$.
For example in the model of BAHN, using $k_I/k_S=1$, we have
$f_{\rm crit}\approx0.38$, and $f_{\rm crit}$
increases for decreasing $k_I/k_S$.
For $k_I/k_S=0$ (no enantiomeric excess)
no bifurcation is possible, i.e.\ $f_{\rm crit}=1$.

The polymerization problem without spatial extent is governed by
the following characteristic time and length scales
\EQ
\tau_{\rm micro}=\left(Q k_S\right)^{-1/2},\quad
\xi_{\rm micro}=\left(Q/k_S\right)^{-1/6},
\EN
where the subscript `micro' is chosen to emphasize that with spatial
extent the problem will be governed by other `macroscopic' time scales
such as $\tau_{\rm diff}$ and $\tau_{\rm turb}$ (see previous section).

Since neither $Q$ nor $k_S$ are well known, it may be advantageous to
assume $\tau_{\rm micro}$ and $\xi_{\rm micro}$ instead:
\EQ
k_S=\xi_{\rm micro}^3\tau_{\rm micro}^{-1},\quad
Q=\xi_{\rm micro}^{-3}\tau_{\rm micro}^{-1}.
\label{ksQ}
\EN
A plausible time scale for polymerization is hours
($\tau_{\rm micro}\sim10^5\s$, say), but it could be much longer,
and typical molecule concentrations are $10^{20}\cm^{-3}$,
but it could also be much less.
The nominal values corresponds to $k_S\sim 10^{-25}\cm^3\s^{-1}$,
$Q\sim 10^{15}\cm^{-3}\s^{-1}$.

The growth rate of instability of the racemic solution, $\lambda$,
scales with $\tau_{\rm micro}^{-1}$, and it is therefore convenient to
introduce the symbol $\lambda_0\equiv\tau_{\rm micro}^{-1}$.
For example, for $k_I/k_S=f=1$, we have $\lambda=0.57\lambda_0$;
see BAHN.

\subsection*{Allowing for spatial extent}

\begin{figure}[t!]
\resizebox{.9\hsize}{!}{\includegraphics{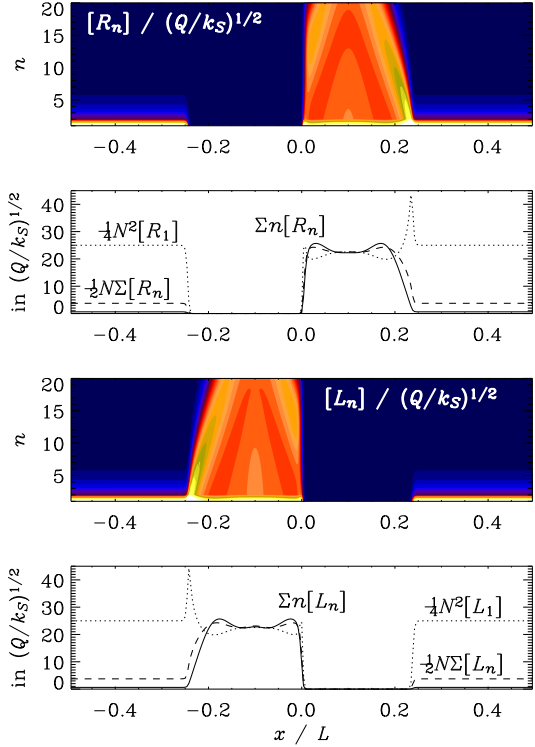}}\caption{
Color/gray scale plots of $[R_n]$ and $[L_n]$ for $t/\tau_{\rm diff}=0.8$
as a function of $x$ and $n$,
and the corresponding dependencies of
$\sum_{n=1}^Nn[R_n]$ and $\sum_{n=1}^Nn[L_n]$ (solid line), compared with
$\half N\sum_{n=1}^N[R_n]$ and $\half N\sum_{n=1}^N[L_n]$ (dashed line), and
$\quarter N^2[R_1]$ and $\quarter N^2[L_1]$ (dotted line), all in units
of $(Q/k_S)^{1/2}$.
The normalized diffusivity is $\kappa/(L^2\lambda_0)=10^{-2}$
and $N=20$.
}\label{pchain}\end{figure}

Before discussing the reduced model and its extension to allow for
spatial extent, let us briefly illustrate the principal effects of
spatial diffusion on the full polymerization model.
If the fidelity is close to unity, an initially small perturbation
to the racemic solution ($[R_n]=[L_n]=2^{-n}[L_1]$ for $k_I=k_S$
and $n\ge2$; see BAHN) grows until a new homochiral state is reached
where there are polymers of only one handedness with equally many of
each length (i.e.\ $[L_n]=2[L_1]$ for $n\ge2$).
The same can happen also locally if we allow for spatial extent
provided the domain is large ($L\gg\xi_{\rm micro}$).
In \Fig{pchain} we show the result of a one-dimensional numerical
calculation in a periodic domain $-L/2<x<L/2$ with $L/\xi_{\rm micro}=10$
and finite diffusivity $\kappa/(L^2\lambda_0)=10^{-2}$, i.e.\
$\tau_{\rm diff}=100\tau_{\rm micro}$, but no flow ($\uu=0$).

Note that the domain divides into regions with opposite handedness.
If one waits a few $e$-folding times ($\lambda^{-1}$), each of the
homochiral sub-domains possesses polymers of appreciable length
(\Fig{pchain}).
At later times the polymers will have grown to maximum length, with a
sharp interface separating domains with right and left handed polymers.
The evolution of these domain walls will be discussed in detail in
the next section using the reduced model of BAHN.

\subsection*{The reduced model}

It turns out that, without changing the basic properties of the model,
a minimal version is still meaningful for $N=2$,
and that the explicit evolution equations for the semi-spoiled polymers,
$[L_2R_1]$ and $[R_2L_1]$, can be ignored.
However, the associated reaction \eq{LnR1react} (and likewise for 
$L\rightleftharpoons R$) still leads to loss terms in all equations
(proportional to $k_I$).
In fact, the very presence of the $k_I$ terms is essential for the
bifurcation toward homochirality.

Thus, we only solve \Eqs{Ln_new3}{Rn_new3} together with \Eqss{L1R1}{lambdaR}.
Following Sandars (2003), we also assume that $C_L=[L_2]$ and $C_R=[R_2]$
(instead of $C_L=E_L$ and $C_R=E_R$, which would yield more complicated
expressions).
A further simplification can be made by regarding $[L_2]$ as a rapidly
adjusting variable that is enslaved to $[L_1]$ (and similarly for $[R_2]$).
This technique is also known as the adiabatic elimination of rapidly
adjusting variables (e.g., Haken, 1983).
\EEqs{Ln_new2}{Rn_new2} become
\EQA
0=k_S[L_1]^2-2[L_2]\Big(k_S[L_1]+k_I[R_1]\Big),\\
0=k_S[R_1]^2-2[R_2]\Big(k_S[R_1]+k_I[L_1]\Big),
\ENA
which are solved for $[L_2]$ and $[R_2]$, respectively.
These quantities
couple back to the equations for $[L_1]$ and $[R_1]$ via $Q_L$ and $Q_R$.
Finally, we also treat the substrate $[S]$ as a rapidly adjusting variable,
i.e.\ we have $k_C[S]=Q/([L_2]+[R_2])$.
We emphasize that the adiabatic elimination does not affect the
accuracy of {\it steady} solutions.
It is convenient to introduce new dimensionless variables,
\begin{equation}
X=[R_1](2k_S/Q)^{1/2},\quad
Y=[L_1](2k_S/Q)^{1/2}.
\end{equation}
and a constant $\lambda_0=(Qk_S/2)^{1/2}$, which has dimensions
of (time)$^{-1}$.
The introduction of the $1/\sqrt{2}$ factor is useful for getting
rid of factors of 2 in the reduced model equations below.

In order to compare first with the SH model we restrict ourselves
to the special case $k_I/k_S=1$, which leads to the revised model
equations
\EQA\label{modeleqn}
\lambda_0^{-1}\dot{X} & = &  {pX^2+qY^2\over\tilde{r}^2}-rX,\\
\lambda_0^{-1}\dot{Y} & = & {pY^2+qX^2\over\tilde{r}^2}-rY,\label{modeleqn:2}
\ENA
where dots denote derivatives with respect to time
and $r=X+Y$, $\tilde{r}^2=X^2+Y^2$, $p=(1+f)/2$ and $q=(1-f)/2$ have 
been introduced for brevity.
The reduced equations resemble those of the SH model
in that both have a quadratic term proportional to $X^2$ (or $Y^2$),
which is quenched either by a $1-r$ factor (in the SH model) or by
a $1/(X^2+Y^2)$ factor in our model.
Furthermore, both models have a backreaction term proportional to $-X$
(or $-Y$), but the coefficient in front of this term
(denoted by $\lambda$ in the SH model) is not constant but equal to $(X+Y)$.

The form of the equations (\ref{modeleqn}) and (\ref{modeleqn:2}) 
suggests a transformation to a new
set of variables
\begin{equation}
{\cal S} = X+Y,\quad {\cal A} = X-Y,
\label{cov}
\end{equation}
in terms of which the enantiomeric excess is in the reduced model simply
\begin{equation}
\eta = {\cal A}/{\cal S}.
\label{zent}
\end{equation}
Adding and subtracting Eqs. (\ref{modeleqn}) and (\ref{modeleqn:2}) from each other leads 
to a new pair of equations in terms of ${\cal S}(t)$ and ${\cal A}(t)$:
\EQA
\lambda_0^{-1}\dot{\cal S} & = & 1-{\cal S}^2\label{bahn2:1},\\
\lambda_0^{-1}\dot{\cal A} & = & 2f {{\cal S}{\cal A}\over {\cal S}^2+{\cal A}^2}-{\cal S}{\cal A}\label{bahn2:2}.
\ENA
Fixed points are: ${\cal S}=1,\, {\cal A}=0,\pm \sqrt{2f-1}$, 
corresponding to the points $((1\pm\sqrt{2f-1})/2,(1\mp\sqrt{2f-1})/2),\, (\frac 12,\frac 12)$ 
in the $(X,Y)$-plane (the ${\cal S}=-1$ fixed points are not 
physical, because $0\leq X,Y\leq1$). 

We see that the enantiomeric excess
is complete only for $f=1$ and that for this case physical
stable points with an enantiomeric excess only appear with $f>1/2$
(for $k_I/k_S=1$).
The growth rate of the enantiomeric excess, $\eta$, from the racemic point, 
$(S,A)=(1,0)$, can be found by linearizing the equations and it 
turns out to be $\dot{\eta}=2f-1$.
For simplicity, we concentrate here on the limiting case $f=1$, 
but also show in a later section, how the results depend on changing $f$ 
(and hence the growth rate).

The first equation, Eq.~(\ref{bahn2:2}), is easily solvable
in terms of hyperbolic functions,
\EQ
{\cal S}(t) = \left\{
\begin{array}{ll}
\coth\lambda_0(t-t_0) & \mbox{for ${\cal S}(0)>1$}, \\
\tanh\lambda_0(t-t_0) & \mbox{for ${\cal S}(0)<1$},
\end{array}\right.
\label{Asoltn}
\EN
where $t_0$ is an integration constant that can also be
expressed in terms of the initial value ${\cal S}(0)$.
From \Eq{Asoltn} we note the long time behavior:
${\cal S}\rightarrow 1$, regardless of the initial value.
Since $0\leq X,Y\leq 1$, we have $0\leq {\cal S}\leq 2$.
The corresponding solution for ${\cal A}$, assuming $f=1$, is
\EQ
{\cal A}(t)={a-\sqrt{a^2+\cosh^2\lambda_0(t-t_0)}
\over\sinh\lambda_0(t-t_0)}
\EN
for ${\cal S}(0)>1$, and
\EQ
{\cal A}(t)={a-\sqrt{a^2+\sinh^2\lambda_0(t-t_0)}
\over\cosh\lambda_0(t-t_0)}
\EN
for ${\cal S}(0)<1$.
Here $a$ is another integration constant that depends on the
initial condition:
\EQ
a=-{\sinh\lambda_0t_0\over2{\cal A}(0)}
\left[{\cal A}(0)^2-{\cal S}(0)^2\right]
\EN
for ${\cal S}(0)>1$, and
\EQ
a=+{\cosh\lambda_0t_0\over2{\cal A}(0)}
\left[{\cal A}(0)^2-{\cal S}(0)^2\right]
\EN
for ${\cal S}(0)<1$.

\subsection*{The reduced model with spatial extent}

As in the model without spatial extent, the polymerization dynamics
seems to be controlled by the dynamics of the shortest polymer.
In other words, the polymerization dynamics of the longer polymers
is enslaved by that of the shorter ones. 

This is the basis of the reduced model of BAHN and hence we
consider a reduced model with spatial extent.
The new set of reduced model equations reads as
\begin{eqnarray}
\begin{array}{l}
\left(\partial_t+\uu\cdot\nab-\kappa\nabla^2\right)
X=\lambda_0F_X(X,Y),\cr
\left(\partial_t+\uu\cdot\nab-\kappa\nabla^2\right)
Y=\lambda_0F_Y(X,Y),
\end{array}
\label{modeleqn2}
\end{eqnarray}
where $F_X$ and $F_Y$ are the right hand sides of the reaction equations
\eq{modeleqn} and \eq{modeleqn:2}.

A crucial property of this model is its bifurcation behavior in time,
as we have seen when considering the models with no spatial extent.
For $\uu=\kappa=0$, we now have the spatially homogeneous solution 
$X(x,t)=Y(x,t)=1/2$, i.e.\ there are equally many right and left 
handed molecules.
Furthermore for $f>1/2$, there are the two additional solutions,
\EQ
X(x,t)=1-Y(x,t)=-1\pm\sqrt{2f-1}.
\EN

The question we wish to answer is how an initially nearly 
homogeneous racemic mixture attains a final state, whether or not it is 
composed of regions of different handedness and what are
the relevant time scales.

\subsection*{The one-dimensional case}
\begin{figure}[ht]
\resizebox{\hsize}{!}{
\includegraphics[width=38mm,angle=0]{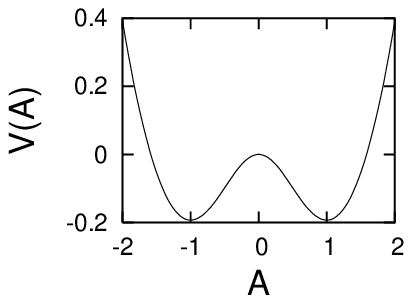}
\includegraphics[width=38mm,angle=0]{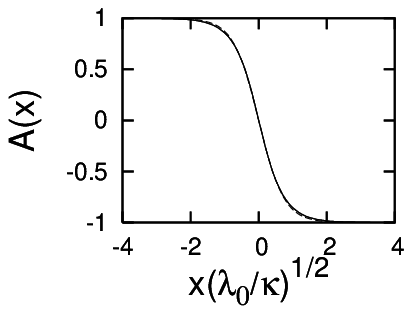}
}\caption{Potential $V({\cal A})$ (for $V_0=0)$ and the solution interpolating
between the left and right handed regions.}
\label{fig2}
\end{figure}

Using the previously introduced variables ${\cal S},\ {\cal A}$ and specializing to 
the one-dimensional case, the equations without advection (\ref{modeleqn2}) take the form
\EQA
\dot{\cal S} & = & \lambda_0(1-{\cal S}^2)+\kappa {\cal S}''\label{bahn4:1}\\
\dot{\cal A} & = & \lambda_0{\cal S}{\cal A}
\left({2\over {\cal S}^2+{\cal A}^2}-1\right)
+\kappa {\cal A}''\label{bahn4:2},
\ENA
where primes indicate a derivative with respect to $x$.
Clearly ${\cal S}(x,t)=1$ is a solution of the first equation. 
Numerically we have seen  that
${\cal S}$ tends to $1$ very quickly from a ${\cal S}\neq 1$ initial state. 
Unless $\kappa$ is very large, one
can view the problem in two parts: first ${\cal S}$ relaxes to 
${\cal S}=1$ and then ${\cal A}$ begins to evolve spatially.
With ${\cal S}=1$, Eq. (\ref{bahn4:2}) takes the form
\EQ
\dot{\cal A} = \lambda_0F_{\cal A}(x,t)+\kappa {\cal A}'',
\EN
where we have introduced
\EQ
F_{\cal A} = {\cal A}\;{1-{\cal A}^2\over 1+{\cal A}^2}.
\label{FA1}
\EN
The time independent equation $\dot{{\cal A}}=0$ can be integrated once and written in
a suggestive form as 
\EQ
{\kappa\over 2\lambda_0}({\cal A}')^2-V({\cal A})=0,
\EN
where 
\EQ
V({\cal A})=\textstyle{\frac 12}{\cal A}^2-\ln(1+{\cal A}^2)+V_0
\EN
is the potential and $V_0$ is an integration constant.
The form of the potential is shown in the left hand panel of \Fig{fig2}.
The minimum is degenerate,
the two minima are located at ${\cal A}=\pm 1$ reflecting the $Z_2$ symmetry of the
theory. The minima correspond to the two stable solutions the system tends to.
As usual, there is a domain wall solution interpolating between the two minima.
The solution is shown in the right hand panel of \Fig{fig2}.
The shape is very well 
approximated by ${\cal A}(x)=\tanh(-\alpha x\sqrt{\lambda_0/\kappa})$, 
where $\alpha\approx 0.58$.
So in one dimension the initially racemic mixture tends to form domain
walls separating left and right handed regions. One can therefore 
see the process as spontaneous symmetry breaking into different domains.
Numerical work confirms the analytical considerations: the region breaks into
different domains that once relaxed to $\eta=\pm 1$, are separated by time 
independent domain walls.
In one dimension, it is hence difficult to achieve a homogeneous mixture unless
the initial perturbation is only directed towards one of the minima.

\begin{figure}[t!]
\resizebox{\hsize}{!}{\includegraphics{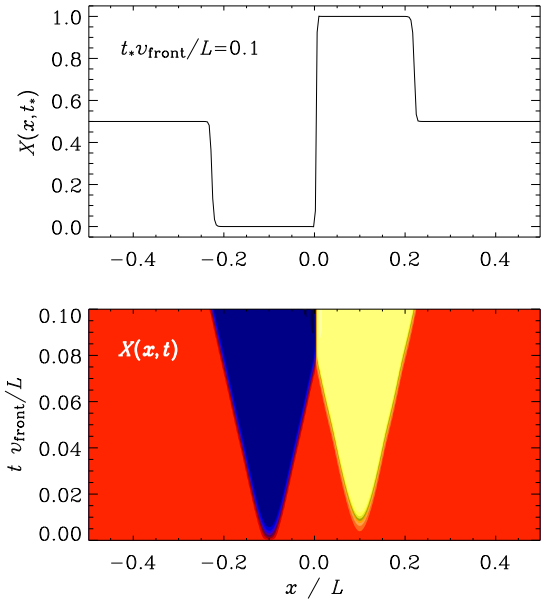}}\caption{
Profile of $X(x,t_*)$ and
space-time diagram of $X(x,t)$ for the one-dimensional problem without
advection and an initial perturbation corresponding to a weak
(amplitude 0.01) right-handed
excess at $x/L=0.1$ (marked in white or yellow) and a somewhat stronger
(amplitude 0.3) left-handed excess at $x/L=-0.1$ (marked in dark or blue).
Note the propagation of fronts with constant speed if the exterior is
racemic (i.e.\ $X=Y=1/2$, shown in medium shades or red)
and a nonpropagating front when
the chirality is opposite on the two sides of the front.
The normalized diffusivity is $\kappa/(L^2\lambda_0)=10^{-2}$,
i.e.\ the same as in \Fig{pchain}.
}\label{1Da}\end{figure}

In \Fig{1Da} we show a
space-time diagram of the solution of the one-dimensional problem 
without advection and an initial perturbation corresponding to a 
weak right-handed excess at $x/L=0.1$ and 
a somewhat stronger left-handed excess at $x/L=-0.1$.
Note the propagation of fronts with constant speed $v_{\rm front}$,
as given by \Eq{vfront}, if the exterior is
racemic (i.e.\ $X=Y=1/2$) and a nonpropagating front when
the chirality is opposite on the two sides of the front.

\subsection*{The higher-dimensional case}
We have seen that in one dimension, the racemic mixture naturally tends
toward a state where there are left and right handed domains. Extending the
analysis is straightforward. Keeping in mind the picture of symmetry breaking
into different ${\cal A}$ minima, one can envision how the spatial distribution 
first breaks up into different domains. The question is then whether these 
domains are stable or not. This question is easily answered by considering a 
spherical bubble, described by
\EQ
\dot{\cal A} = \lambda_0 F_{\cal A}(r,t)+\kappa
\left({{\partial^2}{\cal A}\over {\partial}r^2}+{D-1\over r}
{{\partial}{\cal A}\over {\partial}r}\right).\label{qpr}
\EN
This has a solution where the RHS vanishes, making the solution
stable. However, in more than dimension ($D>1$) this is not the case. 
Consider the time-independent
solution in one dimension: a mechanical analogy is a point mass 
moving under gravity in 
the inverted potential $-V({\cal A})$ with $r$ playing the 
role of time. The solution that interpolates 
between the two minima
corresponds to the point mass rolling down from the top of the hill
and then stopping at the top of the other hill.
This is possible because in the mechanical analogy, the equations
of motion include no friction term. In $D>1$ dimensions, the 
situation is different due to the presence
of the term proportional to $(D-1)\partial_r {\cal A}$.
Analogously to the one-dimensional case,
this term acts as (time dependent) friction.
Hence, there is no spherically symmetric stable solution in more than 
one spatial dimension. The time dependence will work in the direction 
as to shrink the bubble as is evident from Eq.~(\ref{qpr}). In a more 
complicated system, where the wall is not a simple bubble, this analysis 
still describes the behavior qualitatively: the local curvature
determines whether a local region tends to shrink or expand. This 
behavior is also confirmed numerically.

\subsection*{Evolution of the enantiomeric excess}

In order to measure the speed of the process, we need to consider
global quantities that well represent the left or right handedness
of the whole system.
In the reduced model we use the enantiomeric excess, as defined in
\Eq{etadefn}, based on 
$E_R=\int X(\xx,t)\,\dd^D\!x$,
$E_L=\int Y(\xx,t)\,\dd^D\!x$.
Assuming furthermore ${\cal S}=1$, in $D$ dimensions we have
$\eta=\int {\cal A}\,\dd^D\!x$,
and so for $f=1$ we have
\EQ
{d\eta\over dt}={\lambda_0\over L^D}\int F_{\cal A}(\xx,t)\,\dd^D\!x,
\EN
because the integral over the laplacian vanishes
for periodic boundary conditions.
Here, $L$ is the size of the domain and $D=2$ or 3 for
two and three-dimensional systems, respectively.

\begin{figure}[t!]
\resizebox{\hsize}{!}{\includegraphics{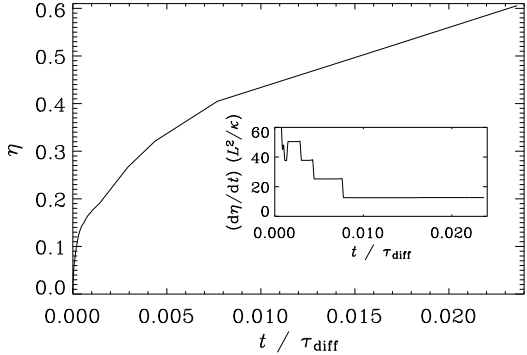}}\caption{
Evolution of $\eta$ for a two-dimensional simulation
with $\kappa/(L^2\lambda_0)=2\times10^{-4}$,
$L/\xi_{\rm micro}=16\pi$, and $1024^2$ mesh points resolution.
The inset shows the normalized slope.
Note the fours distinct regimes with progressively decreasing slope.
}\label{peta}\end{figure}

We have calculated the evolution of $X(\xx,t)$ and $Y(\xx,t)$ through
numerical simulations in two and three spatial dimension,
with and without advection.
We use the {\sc Pencil Code}\footnote{
\url{http://www.nordita.dk/data/brandenb/pencil-code}}, which is a modular
high-order finite-difference code (sixth order in space and third order
in time) for solving general partial differential equations and,
in particular, the compressible hydrodynamic equations.
We usually start with random initial conditions where $X$ and $Y$ are
uniformly distributed between 0 and 1.
The results of the simulations
are shown in \Figss{peta}{pres_diffadv}.

\begin{figure}[t!]
\resizebox{\hsize}{!}{\includegraphics{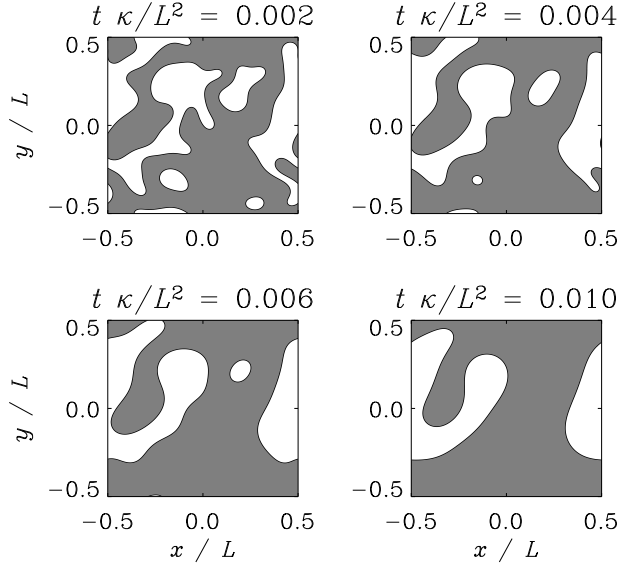}}\caption{
Plot of $X(x,y)$ at four different times for the same simulation as in
\Fig{peta}.
White indicates $X=1$ and dark indicates $X=0$.
Note that the number of disconnected regions decreases from 4 in the
first plot to 3, 2, and 1.
}\label{pxxx_chiral}\end{figure}

In \Fig{peta} we show the evolution of $\eta$ for a model with large spatial
extent; $L/\xi_{\rm micro}=16\pi\approx50$.
Note that the evolution proceeds in steps with piecewise constant values
of $\dot{\eta}$ (see the inset).
The progressively decreasing value of $\dot{\eta}$ is associated with
a reduction of the number ${\cal N}$ of topologically connected regions
in the full domain.
Looking at \Fig{pxxx_chiral}, ${\cal N}$ decreases from
${\cal N}=4$ at $t/\tau_{\rm diff}=0.002$ to
${\cal N}=1$ at $t/\tau_{\rm diff}=0.01$, in accord with the change
of slope of $\eta(t)$ see in \Fig{peta}.

\begin{figure}[t!]
\resizebox{\hsize}{!}{\includegraphics{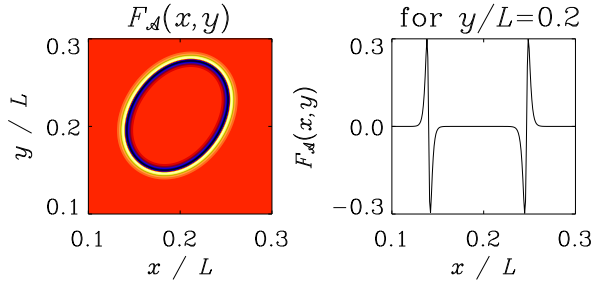}}\caption{
Color/gray scale representation of the quantity 
$F_{\cal A}(x,y,t)$ corresponding to the third plot
in \Fig{pxxx_chiral} near the location of the small isolated ring.
Note the sharp alternations of the sign of $F_{\cal A}$ just at the boundary
of the ring.
}\label{pfac}\end{figure}

A blow-up of the little white region in the third panel
of \Fig{pxxx_chiral} ($t/\tau_{\rm diff}=0.006$) is shown in \Fig{pfac}.
Here, instead of $X$, we show $F_{\cal A}$ along with a cross-section of the loop.
Both inside and outside the ring $F_{\cal A}=0$, and $F_{\cal A}\neq0$
only at the location of the ring-like front.

While the fronts shrink and finally disappear, they appear surprisingly robust and
the relaxation to the final homochiral state is a slow process compared to the initial
local relaxation to many disconnected regions with $\eta=\pm 1$.
So, even though the local relaxation
occurs on the time scale $\tau_{\rm micro}$,
the total homochiralization process takes place on the much longer time
scale, $\tau_{\rm diff}$.

The piecewise linear scaling of $\eta(t)$ is associated with the simple
front structure shown in \Fig{pfac}.
We recall that $\dot{\eta}$ is directly proportional to $F_{\cal A}$.
However, not only is $F_{\cal A}=0$ outside the front, the positive and
negative contributions on either side of the front {\it almost} cancel.
In fact, if the front were straight, the two contributions
from either side of the front would cancel exactly, in which case
$\dot{\eta}=0$.
The only reason the two contributions do not cancel exactly is because the
inner and outer parts of the front have different radii, $r_1$ and $r_2$.
Denoting the width of each ring by $\Delta r$, and since their separation
is also $\approx\Delta r$, we have
$\dot{\eta}\propto (r_2-r_1)\Delta r\approx(\Delta r)^2$.
Note that $(\Delta r)^2$ is independent of time.
Thus, the value of $\dot{\eta}$ is independent of how complicated the
structures are, as long as its topology is the same.
Furthermore, our simulations confirm that $(\Delta r)^2$ is proportional
to $\kappa/\lambda_0$ (see below for a quantitative demonstration).

\begin{figure}[t!]
\resizebox{\hsize}{!}{\includegraphics{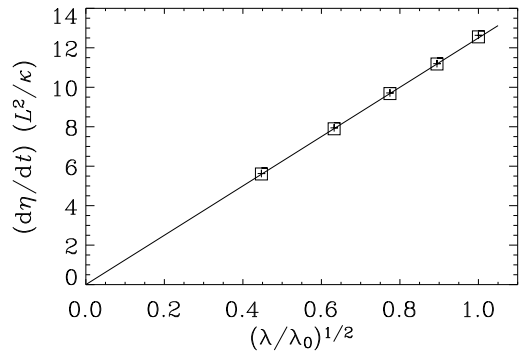}}\caption{
Normalized values of $\dot\eta$ as a function of $\lambda$ for
$\kappa/(L^2\lambda_0)=6.3\times10^{-6}$ (diamonds) and
$1.3\times10^{-5}$ (plus signs).
Both groups of data points overlap and match the fit formula \Eq{fitformula}.
In order to reduce noise in the date we have chosen a single ring-like
structure as initial condition.
}\label{pres_difflamf}\end{figure}

So far we have always assumed that $k_I/k_S=f=1$.
If this is not the case, then $\lambda/\lambda_0$ ($=2f-1$) will be
different from unity, and so will be $|{\cal A}|$.
Instead, ${\cal A}=\pm\sqrt{\lambda}$ [see Eqs.~(43) and (44) of BAHN].
Furthermore, the expression for $F_{\cal A}$ is different from that
given in \Eq{FA1}, and is now
\EQ
F_{\cal A} = {\cal A}\;{\lambda-{\cal A}^2\over 1+{\cal A}^2}.
\label{FA2}
\EN
However, as before, $F_{\cal A}=0$ on either side of the front, and a
finite contribution comes only from an imperfect cancellation
from within the front, where it is governed by the value of
${\cal A}\propto\sqrt{\lambda}$.
This is also confirmed numerically and we find (see \Fig{pres_difflamf})
\EQ
{\dd\eta\over\dd t}\approx12.5\,{\cal N}\,
\left({2\lambda^2\over Q k_s}\right)^{1/4}{\kappa\over L^2},
\label{fitformula}
\EN
where we have included the dependence on the number of
topologically disconnected structures ${\cal N}$, and
the growth rate $\lambda_0$ has been expressed in terms of $Q$ and $k_S$.
Note, in particular, that there is no evidence for a correction arising
possibly from the denominator of \Eq{FA2}.

\begin{figure}[t!] 
\resizebox{\hsize}{!}{\includegraphics{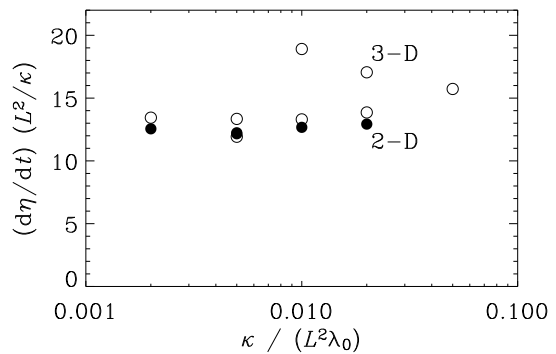}}\caption{
Normalized values of $\dd\eta/\dd t$ obtained from simulations in
two dimensions (dots) and three dimensions (circles).
In the three-dimensional runs, the resolution is varied between
$32^3$ for $\kappa/(L^2\lambda_0)=8\times10^{-3}$ and
$256^3$ for $\kappa/(L^2\lambda_0)=3\times10^{-4}$.
In all cases $L/\xi_{\rm micro}=2\pi$.
}\label{pres_diff}\end{figure}

In three dimensions, the evolution of $\eta(t)$ is no longer perfectly
linear, but an average value of $\dot{\eta}$ can still be determined.
It turns out that these values are rather similar to the
two-dimensional case; see \Fig{pres_diff}.
In three dimensions the only finite contribution to $F_{\cal A}$
comes from a pair of shells with radii $r_1$ and $r_2$, so
$\dot{\eta}\propto (r_2^2-r_1^2)\Delta r\approx2\overline{r}\,(\Delta r)^2$,
where $\overline{r}$ is some mean radius that depends on time as the
shells shrink.
This explains the departures from linear scaling in the three-dimensional
case.
In terms of $\kappa$ one has $\dot{\eta}\propto\kappa\,\overline{r}/L^3$.
The observed similarity between two and three-dimensional cases suggests
that $2\overline{r}/L=O(1)$.

The main conclusion from these studies is that
the growth of the enantiomeric excess occurs via diffusion and not
via front propagation.

\begin{figure}[t!]
\resizebox{\hsize}{!}{\includegraphics{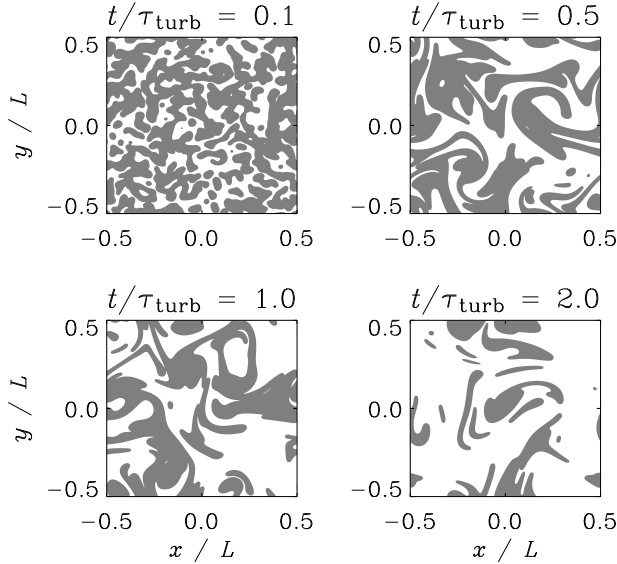}}\caption{
Plot of $X(x,y,t)$ in a two-dimensional simulation with turbulence.
As in \Fig{pxxx_chiral}, $\kappa/(L^2\lambda_0)=2\times10^{-4}$,
$L/\xi_{\rm micro}=16\pi$, and $1024^2$ mesh points resolution.
Note the narrow elongated structures enhancing significantly the chance of
extinguishing one or the other life form.
}\label{turb}\end{figure}

\begin{figure}[t!]
\resizebox{\hsize}{!}{\includegraphics{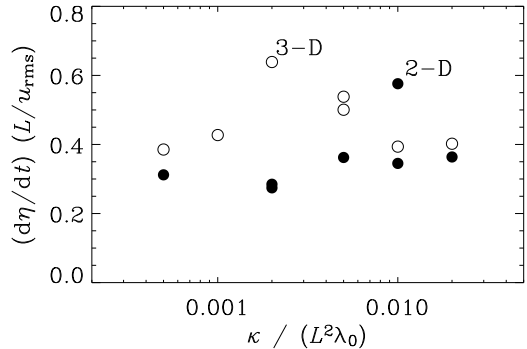}}\caption{
Normalized values of $\dd\eta/\dd t$ obtained from simulations with
turbulence in two dimensions (dots) and three dimensions (circles).
In the three-dimensional runs, the resolution is varied between
$32^3$ for $\kappa/(L^2\lambda_0)=3\times10^{-3}$ and
$256^3$ for $\kappa/(L^2\lambda_0)=8\times10^{-5}$.
In all cases $L/\xi_{\rm micro}=2\pi$.
}\label{pres_diffadv}\end{figure}

Finally we turn to the case of a finite advection velocity $\uu$.
The evolution equations for $X$ and $Y$ are now supplemented by an
evolution equation for the three components of $\uu$ and, in this case,
for the logarithmic density.
The turbulence is forced in a narrow band of wavenumbers that are
about 3 times the minimum wavenumber of the box.
For the present discussion the important quantities are the root mean square
velocity, $u_{\rm rms}$, and the
energy carrying scale $\ell$ ($=L/3$ in the present case).
The details of such simulations\footnote{
Animations of both turbulent and purely diffusive solutions can be found
at \url{http://www.nordita.dk/~brandenb/movies/chiral}}
have been discussed elsewhere in
more detail (e.g., Brandenburg, 2001; Brandenburg \ea 2004b).

In the presence of turbulence, originally detached regions with growing
right and left-handed polymers are driven into close contact, enhancing
thereby the probability of extinguishing one of the two possible file
forms; see \Fig{turb}.
It turns out that $\dot\eta$ still grows roughly linearly in time,
but now at a dramatically enhanced rate; see \Fig{pres_diffadv}.
As expected, there is no dependence on $\kappa$
and the physically relevant quantities are instead $u_{\rm rms}$ and $\ell$.

It is possible -- and indeed customary -- to express the enhanced turbulent
transport in terms of a turbulent (eddy) diffusivity.
Combining $\dot\eta\approx(0.3...0.6)\,u_{\rm rms}/L$ (from \Fig{pres_diffadv})
and $\dot\eta\approx12.5\kappa_{\rm turb}/L^2$ [the latter being as a definition of
$\kappa_{\rm turb}$; cf.\ \Eq{fitformula}], we find
\EQ
\kappa_{\rm turb}\approx(0.08...0.15)\,u_{\rm rms}\ell,
\EN
where we have used $\ell=L/3$.
This is somewhat below the canonical estimate of $\onethird u_{\rm rms}\ell$
(Prandtl 1925) for turbulent diffusion of a passive scalar,
but consistent with results of simulations of passive scalar diffusion
(e.g., Brandenburg \ea 2004b).
Thus, the fact that the flow has reactive properties does not seem to
affect the turbulent mixing.

It should also be pointed out that the overall structure of $X$ and $Y$ is now
rather irregular and no longer front-like as in the case of pure diffusion.
Thus, the fact that we can define a turbulent value of the diffusivity
does not mean that with turbulence the mean distributions of $X$ and $Y$ can
accurately be described by a diffusion term.

\subsection*{Discussion}

Our studies have shown that, without advection, the enantiomeric excess grows
toward homochirality with the speed 
$\dot{\eta}\sim\kappa/L^2$, where $L$ is the physical size of the system;
see \Eq{fitformula}.
This can be much slower than the propagation of epidemics which
proceeds on the faster time scale $\tau_{\rm front}=L/v_{\rm front}$.
For the Earth, this would yield a time scale for homochiralization of
$\tau\sim 10^{17}\s$, i.e.\ comparable to the age of the universe. 
So the diffusive process is too slow
in creating large homogeneous areas of one chirality,
even though within a local region, a single
handed state is created on the time scale $\lambda_0^{-1}$, which could
be on the order of days [see the discussion after \Eq{ksQ}], or even
years (as in the context of the spreading of epidemics).

Turbulence, however, changes the situation drastically:
the speed toward homochirality is effectively 
$\dot{\eta}\sim u_{\rm rms}/L$, which, assuming a root mean square
flow speed of say $1\cms$, 
leads to time scales of a hundred years, in rough agreement
with results from ocean mixing studies where
horizontal eddy diffusivity is found to be of the order of 
$10^3\mtwopers$ (Stewart 2003, Chap.~8.5). 

Comparisons to present day oceanographic studies must be done with care, however, as
there are a number of factors that can affect both horizontal and vertical mixing such
as the higher salinity in the early oceans (Knauth 1998)
and most importantly the possible
lack of continents on the early Earth. The role of continents is manifold as 
they aid in the vertical mixing and are possibly needed in creating
large scale flows (Garrett, 2003).
Hence, the conditions and mixing in the early oceans may have
been quite different from the complex system of currents what we observe 
today.
In addition, frequent impacts on the early Earth may not only have
led to increased mixing, but also to global extinction of life more than
once (Wilde \ea 2001).

Although it is impossible to know with any degree of certainty the amount
of mixing in the early ocean, the time scales may have been rather long.
If this is the case, the question of homochirality would remain open.
Thus, it might therefore be interesting to investigate in more detail whether
the chirality of biomarkers from early marine sediments can be determined.
An obvious difficulty with this proposal is the fact that dead matter tends
to racemize spontaneously on time scales that are typically shorter than
a million years (e.g., Bada, 1995).

\section*{Acknowledgments}

The Danish Center for Scientific Computing is acknowledged
for granting time on the Linux cluster in Odense (Horseshoe).

\section*{References}

\begin{list}{}{\leftmargin 3em \itemindent -3em\listparindent \itemindent
\itemsep 0pt \parsep 1pt}\item[]

Bada, J. L.\ynat{1995}{374}{594}
{595}{Origins of homochirality}

Brandenburg, A.\yapj{2001}{550}{824}
{840}{The inverse cascade and nonlinear alpha-effect in simulations
of isotropic helical hydromagnetic turbulence}

Brandenburg, A., Andersen, A., Nilsson, M., and H\"ofner, S.\poleb{2004a}
{Homochiral growth through enantiomeric cross-inhibition}
Preprints available online at: \url{http://arXiv.org/abs/q-bio/0401036} (BAHN).

Brandenburg, A., K\"apyl\"a, P., and Mohammed, A.\ypf{2004b}{16}{1020}
{1027}{Non-Fickian diffusion and tau-approximation from numerical turbulence}

Cleaves, H. J. and Chalmers, J. H.\yab{2004}{4}{1}
{9}{Extremophiles may be irrelevant to the origin of life}

Collins, L. J., Poole, A. M., and Penny, D.\yjour{2003}{Appl.\ Bioinform.}{2}{S85}
{S95}{Using ancestral sequences to uncover potential gene homologues}

Cotterill, R.\ybook{2002}{Biophysics: an introduction}
{J. Wiley \& Sons, Chichester}

Fisher, R. A.\yjour{1937}{Ann.\ Eugenics}{7}{353}
{369}{The wave of advance of advantageous genes}

Frank, F.\yjour{1953}{Biochim.\ Biophys.\ Acta}{11}{459}
{464}{On Spontaneous Asymmetric Synthesis}

Garrett, C.\ysci{2003}{301}{1858}{1859}{Internal tides and ocean mixing}

Haken, H.\ybook{1983}{Synergetics -- An Introduction}{Springer: Berlin}

Joyce, G. F., Visser, G. M., van Boeckel, C. A. A., van Boom, J. H.,
Orgel, L. E., and Westrenen, J.\ynat{1984}{310}{602}
{603}{Chiral selection in poly(C)-directed synthesis of oligo(G)}

Kaliappan, P.\yphy{1984}{D 11}{368}
{374}{An exact solution for travelling waves of $u_t=Du_{xx}+u-u^k$}

Knauth, L. P.\ynat{1998}{395}{554}
{555}{Salinity history of the Earth's early ocean}

Kozlov, I. A., Pitsch, S., and Orgel, L. E.\ypnas{1998}{95}{13448}
{13452}{Oligomerization of activated D- and L-guanosine mononucleotides
on templates containing D- and L-deoxycytidylate residues}

Martin, W. and Russell, M. J.\yptrs{2003}{B 358}{59}
{83}{On the origins of cells: a hypothesis for the evolutionary
transitions from abiotic geochemistry to chemoautotrophic prokaryotes,
and from prokaryotes to nucleated cells}

Martinez, J. S., Zhang, G. P., Holt, P. D., Jung, H.-T., Carrano, C. J.,
Haygood, M. G., Butler, A.\ysci{2000}{287}{1245}
{1247}{Self-assembling amphiphilic siderophores from marine bacteria}

M\'endez, V., Fort, J., Rotstein, H. G., and Fedotov, S.\ypreN{2003}{68}{041105}
{Speed of reaction-diffusion fronts in spatially heterogeneous media}

Milner-White, J. E. and Russell, M. J.\poleb{2004}
{Sites for phosphates and iron-sulfur thiolates in the first membranes:
3 to 6 residue anion-binding motifs (nests)}

Mojzsis, S. J., Arrhenius, G., McKeegan, K. D., Harrison, T. M.,
Nutman, A. P., Friend, C. R. L.\ynat{1996}{384}{55}
{59}{Evidence for life on Earth before 3,800 million years ago}

Murray, J. D.\ybook{1993}{Mathematical Biology}{Springer, Berlin}

Nelson, K. E., Levy, M., and Miller, S. L.\yjour{2000}
{Proc.\ Nat.\ Acad.\ Sci.\ U.S.A.}{97}{3868}
{3871}{Peptide Nucleic Acids rather than RNA may have been
the first genetic molecule}

Nielsen, P. E.\yoleb{1993}{23}{323}
{327}{Peptide nucleic acid (PNA): A model structure for the primordial
genetic material}

Pal, D., S\"uhnel, J., and Weiss,
M. S.\yjourS{2002}{Angew.\ Chem.\ Int.\ Ed.}{41}{4663}
{4665}{New principles of protein structures: nests, eggs--and what next?}

Pooga, M., Land, T., Bartfai, T., Langel, \"U.\yjour{2001}{Biomol.\ Eng.}{17}{183}
{192}{PNA oligomers as tools for specific modulation of gene expression}

Prandtl, L.\yjour{1925}{Zeitschr.\ Angewandt.\ Math.\ Mech.}{5}{136}
{139}{Bericht \"uber Untersuchungen zur ausgebildeten Turbulenz}

Rasmussen, S., Chen, L., Nilsson, M., and
Abe, S.\yjour{2003}{Artif.\ Life}{9}{269}
{316}{Bridging nonliving and living matter}

Russell, M. J. and Hall, A. J.\yjour{1997}{J.\ Geol.\ Soc.\ Lond.}{154}{377}
{402}{The emergence of life from iron monosulphide bubbles at a submarine
hydrothermal redox and pH front}

Sandars, P. G. H.\yoleb{2003}{33}{575}
{587}{A toy model for the generation of homochirality during polymerization}

Saito, Y. and Hyuga, H.\yjour{2004a}{J.\ Phys.\ Soc.\ Jap.}{73}{33}
{35}{Complete homochirality induced by the nonlinear autocatalysis
and recycling} (SH)

Saito, Y. and Hyuga, H.\sjour{2004b}{J.\ Phys.\ Soc.\ Jap.}
{Homochirality proliferation in space}
Preprints available online at: \url{http://arXiv.org/abs/physics/0405068}.

Schopf, J. W.\ysci{1993}{260}{640}
{646}{Microfossils of the early archean apex chert -- new evidence of
the antiquity of life}

Stewart, R. H.\ybook{2003}{Introduction to Physical Oceanography}
{\url{http://oceanworld.tamu.edu/resources/ocng_textbook/contents.html}}

Tedeschi, T., Corradini, R., Marchelli, R., Pushl, A., and 
Nielsen, P. E.\yjour{2002}{Tetrahedron: Asymmetry}{13}{1629}
{1636}{Racemization of chiral PNAs during solid-phase synthesis: effect
of the coupling conditions on enantiomeric purity}

Ward, P. D. and Brownlee, D.\ybook{2000}{Rare Earth: why complex life is
uncommon in the Universe}{Copernicus, Springer: New York}

Watson, J. D. and Milner-White, E. J.\yjour{2002}{J.\ Mol.\ Biol.}{315}{187}
{198}{A novel main-chain anion-binding site in proteins: the nest. A
particular combination of $\phi,\psi$ values in successive residues gives
rise to anion-binding sites that occur commonly and are found often at
functionally important regions}

Wattis, J. A. D. and Coveney, P. V. (2004) Symmetry-breaking in
chiral polymerization, {\em Orig.\ Life Evol.\ Biosph.} (in press).
Preprints available online at: \url{http://arXiv.org/abs/physics/0402091}.

Wilde, S. A., Valley, J. W., Peck, W. H., and
Graham, C. M.\ynat{2001}{409}{175}
{178}{Evidence from detrital zircons for the existence of continental crust
and oceans on the Earth 4.4 Gyr ago}

\end{list}

\vfill\bigskip\noindent{\it
$ $Id: paper.tex,v 1.90 2004/07/05 20:08:30 brandenb Exp $ $}

\end{document}